\documentclass[sigconf]{acmart}

\AtBeginDocument{%
  \providecommand\BibTeX{{%
    \normalfont B\kern-0.5em{\scshape i\kern-0.25em b}\kern-0.8em\TeX}}}

\setcopyright{skcopy}
\usepackage{hyperref}
\hypersetup{
    colorlinks=true,
    linkcolor=blue,
    filecolor=magenta,      
    urlcolor=blue,
}

\newcommand{\ra}{\operatorname{\rangle}}
\newcommand{\la}{\operatorname{\langle}}

\newcommand{\calA}{{\mathcal{A}}}
\newcommand{\calR}{{\mathcal{R}}}
\newcommand{\calI}{{\mathcal{I}}}
\newcommand{\calM}{{\mathcal{M}}}
\newcommand{\certus}{{\sf Certus}}
\newcommand{\smallpara}[1]{\vspace{.5ex} \noindent {\bf #1}}
\begin{document}

\title{Linking Graph Entities with Multiplicity and Provenance}


\author{Jixue Liu}
\orcid{0000-0003-4014-4976}
\affiliation{%
  \institution{University of South Australia}
  %\streetaddress{P.O. Box 1212}
  \city{Adelaide}
  \state{Australia}
  %\postcode{SA 5095}
} 
\email{jixue.liu@unisa.edu.au}
\orcid{0000-0003-4014-4976}

\author{Selasi Kwashie}
\authornote{Corresponding author.}
\affiliation{%
	\institution{Data61 - CSIRO}
	\streetaddress{Gate 13, Kintore Avenue}
	\city{Adelaide}
	\country{Australia}}
\email{selasi.kwashie@data61.csiro.au}
\orcid{0000-0003-4014-4976}

\author{Jiuyong Li}
\affiliation{%
  \institution{University of South Australia}
  %\streetaddress{P.O. Box 1212}
  \city{Adelaide}
  \state{Australia}
  %\postcode{SA 5095}
}
\email{jiuyong.li@unisa.edu.au}

\author{Lin Liu}
\affiliation{%
  \institution{University of South Australia}
  %\streetaddress{P.O. Box 1212}
  \city{Adelaide}
  \state{Australia}
  %\postcode{SA 5095}
}
\email{lin.liu@unisa.edu.au}

\author{Michael Bewong}
\affiliation{%
  \institution{Charles Sturt University}
  \city{Wagga Wagga}
  \country{Australia}
}
 \email{mbewong@csu.edu.au}

\renewcommand{\shortauthors}{J. Liu, et al.}

\begin{abstract}
Entity linking and resolution is a fundamental database problem with 
applications in data integration, data cleansing, information retrieval, 
knowledge fusion, and knowledge-base population.
It is the task of accurately identifying multiple, differing, and possibly 
contradicting representations of the same real-world entity in data. 
In this work, we propose an entity linking and resolution system capable of linking 
entities across different databases and mentioned-entities extracted 
from text data. Our entity linking/resolution solution, called \certus, uses 
a graph model to represent the profiles of entities. The graph model is 
versatile, thus, it is capable of handling multiple values for an attribute 
or a relationship, as well as the provenance descriptions of the values.
Provenance descriptions of a value provide the settings of the value, such 
as validity periods, sources, security requirements, etc. 
This paper presents the architecture for the entity linking system, the 
logical, physical, and indexing models used in the system, and the general 
linking process. 
Furthermore, we demonstrate the performance of update operations of the 
physical storage models when the system is implemented in two state-of-the-art 
database management systems, HBase and Postgres. 
\end{abstract}





\maketitle

\section{Introduction}
In entity linking and resolution, {\em entities} refer to real-world objects 
(e.g., people, locations, vehicles, etc.) and real-world happenings (e.g., 
events, meetings, interactions, etc.). Entities are described by data in 
information systems. However, the descriptions may be repeated and different 
in these systems. In a database, for instance, a person may have more than one 
record in a table, and the records may have repeating, differing and contradicting 
information about the person. Likewise, two different databases may capture 
different information about the same entity. For example, a medical database only 
concerns with a person's health related properties, whereas an immigration database 
only concerns with the truthfulness of a person's identity. 

A description of an entity is called a {\em profile}, and it can be a record 
in a relational database or a paragraph of words about an entity in a document. 
An entity may have multiple profiles in one or more sources. 
In other words, multiple profiles in one or more databases (or documents) 
may refer to the same real-world entity.  

Once the profiles of entities are captured into a database, the profiles and the 
entities become separated in that the users of the database know the profiles, 
but possibly, not the entities. This separation raises a serious issue. Answering 
the question of whether a  given profile refers to a particular real-world entity 
is non-trivial and challenging. For example, given the profile: {\tt\{name: Michael 
Jordan,  nationality: American,  occupation: athlete\}},  there are at least 
four real-world persons whose profiles in Wikipedia match this  description 
(see~\cite{wiki-mike} for details). A dual problem to the above problem is whether 
two profiles, which may look similar or very different, refer to the same real-world 
entity. This dual problem is as hard as the above problem. 

The goal of entity linking research is to design methods to derive an answer to 
the dual question: {\em do a pair of given profiles refer to the same entity}? 
When a pair of profiles are found to refer to the same entity, one of two actions 
may be  taken. One is to remove of one of the profiles. This is called {\em 
deduplication}. The other is to merge the two profiles and this is called {\em 
resolution/linking}\footnote{We consider linking and resolution, and use the terms
interchangeably in this work}. 

Three complications make the linking/deduplication task more difficult. 
The first complication is from  {\em non-alignment of attributes and relationships}. 
That is, different profiles describe entities using different attributes and/or
relations.  
This is illustrated by the profiles $p_1$ and $p_2$ in Table \ref{tb:prof_complex}.
The two profiles have different attributes except for the {\tt name} attribute.
The non-aligned attributes make their match less possible. 
The second complication is from the {\em multiplicity of values}. 
For example, compared with the profile $p_1$, the profile $p_3$ has two {\tt name} 
values. The third complication is the presence of {\em provenance data}. Provenance 
data describes the background information of a value as well as the 
validity period(s), security \& access restriction(s), source(s), etc., of a value. 
For example, in $p_4$,  {\tt\{since 2005\}} specifies when the {\tt name} 
`{\tt George}' started being used, and {\tt\{2010\}} indicates when the {\tt height}
valued `{\tt 160}' was taken. 
Unlike non-alignment, multiplicity and provenance of values can be useful as 
they provide more information. However, their usefulness comes at a cost:
they require more powerful matching algorithms and data structures to enable 
effective usage. 

%\\
\begin{table}
\caption{Complexity of entity profiles }
\label{tb:prof_complex}
\vspace{-6pt}
\small 
\begin{tabular}{c|l}\hline 
   $p_1$ & {\tt\{name:George, birth-date:1/Jan/2000\}} \\
   $p_2$ & {\tt \{name:George, sex:male, height:160\} }\\
   $p_3$ & {\tt \{name:George, name:Jord, sex:male, height:160\}} \\
   $p_4$ & {\tt\{name:George \{since 2005\}, name:Jord, height:160\{2010\}\}}\\
   \hline
\end{tabular}
\vspace{-12pt}
\end{table}

This paper presents the system supporting our entity linking method \certus ~\cite{KwashieLLLSY19}, the data and index models that enable multiplicity and
provenance of attribute and relationship values to be accurately captured and 
leveraged for effective and efficient entity linking.
The contributions of the paper are as follows. 

\begin{itemize}
    \item First, we present the architecture of our entity linking system (Section~\ref{sec-arch}). This architecture enables textual data to be 
    processed and the entities described in the texts can be linked to entity
    profiles from other data sources. The architecture uses Elasticsearch\footnote{https://www.elastic.co/}, 
    an index engine, to increase the linking and search efficiencies.
    \item Secondly, we propose a graph model for entity linking involving multiplicity of attribute and relation values with provenance information (subsection~\ref{sec-model-prof}).
    In this model, the attributes and relations of profiles are well-represented by
    lists of sets (of attribute/relation, value, and provenance), instead of dictionaries of attribute- and relation-value pairs. 
    Our model enables provenance and value-multiplicity to be captured, indexed and used
    correctly.
    \item Thirdly, we propose physical models for the storage of the graph 
    of entity profiles; detail the index structures that
    support effective search and blocking operations (subsections~\ref{sec-phy-model} \&~\ref{sec-index}); 
    and give the processes in the entity linking component of the system (Section~\ref{sec-linking}).
    \item Lastly, we show experiments about the time performances of our 
    physical model implementations on both relational and non-relational 
    database management systems (Section~\ref{sec-exp}).  
\end{itemize}

\section{Related work}

Entity linking and resolution is a well-known database problem that has attracted volumes 
of research in the literature, especially in the relational data setting. 
% Methods such as distance-based, probabilistic weighting, blocking, supervised 
% and unsupervised methods have been proposed in the literature. 
Readers are referred to~\cite{syn-Naumann} for details. 
In general, the existing works focus on two main directions: accuracy and efficiency. 
The accuracy concern is on finding true matches of different entity profiles when they refer 
to the same real-world entity without introducing false matches. A more specific term called 
{\it efficacy} is defined to mean accuracy in \cite{record-linkage-survey2006}. 
The efficiency issue is about alleviating the infeasible pairwise comparison of profiles, 
and making the linking process scalable in large data.

For accurate entity linking and deduplication, early works on the subject 
examined many methods such as cosine similarity match, distance-based match, TF/IDF, and 
Soundex. The well known similarity measures for entity linking are summarized and reviewed in~\cite{survey-similarity-meas-algo-sigmod06}; and the work in~\cite{evalu-entity-reso-approaches-2010vldb} presents a comparative evaluation of some 
existing works. 

%One well known problem for entity linking is efficiency. 
The efficiency problem has also drawn significant research attention. 
The complexity of calculating the exact similarity between profile pairs is $O(n^2)$. 
Given a large number of entity profiles, say $n=100$ million, the time for computing similarity is too long to be practical.
Thus, several ideas have been introduced in the literature to address the problem, like 
canopy (sorting and moving window), hierarchical, bucketing (clustering), and 
indexing approaches. 
In practice, the indexing approach has been found to be more useful, resulting in the proposal 
of a plethora of indexing methods in the literature (see~\cite{comp-approx-blocking-ER-2016vldb,survey-indexing-tech-2012} for surveys of techniques).

In the recent years, there has been an increasing research interest in linking entity-mentions in texts to existing entities in knowledge-bases. From Wiki Miner 
in \cite{Wikipedia-miner-2008cikm}, many works have been produced in this area and 
are reviewed in~\cite{survey-entity-kb-shen2015,survey-entity-mention-2018}. 
The fundamental steps in text-based linking include: entity-mention detection, 
candidate matching-entity generation, and candidate matching-entity ranking. 
The work in~\cite{survey-entity-mention-2018} reviews the methods for detecting
entity-mentions in texts. Whereas the review paper~\cite{survey-entity-kb-shen2015}
summarizes the details of how features (such as the mentions, types, contexts, etc.)
and models
(e.g., unsupervised, supervised, probabilistic, graph-based, and combined methods) 
are used in the ranking of candidate matching-entities. 
The efforts toward ranking is continuing, and the work in~\cite{relation-as-latent-NN-Edinb-2018ACL} aims to identify effective 
relationship words among entity-mentions to increase the accuracy of linking. 

Most data management and software companies claim to support entity linking in
structured data, but the systems are often not available for evaluation. 
In contrast, a number of open source research frameworks are available on 
entity linking in text data. For example, \cite{Wikipedia-miner-2008cikm} 
proposes a method to extend terms in texts using Wikipedia pages.
\cite{tagme-10} is a framework tagging terms in {\it short} texts by Wikipedia 
pages, which is then followed by the works in~\cite{fr-tagme-to-WAT-2014,reproduce-tagme}
for software improvement. 
\cite{fr-tagme-to-WAT-2014} and \cite{dexter-mention2wiki-2013} are other tools 
that contain a three step implementation for linking entity-mentions in text to 
Wikipedia pages. The work in \cite{gerbil-benchmk-framework-15WWW} sets up a 
framework for entity linking work to be tested and evaluated.  

There exists works in the literature on the support and use of provenance for entity 
linking. For example, \cite{provenance-model-er} is on provenance modeling and capture 
for entity linking whereas \cite{provenance-er} presents a provenance-aware framework 
for improving entity linking results. Our work models, supports, and leverages provenance 
as well as attribute- and relation-value multiplicity for accurate entity linking in 
both structured and text data.

% This paper presents the architectural details of the entity linking solution proposed
% in~\cite{KwashieLLLSY19}. In this work, we detail the system architecture, the data models and 
% indexing structures that support our entity linking system, named \certus. 
% We present the physical implementations for our profile-pair similarity storage model in both
% relational and non-relational database management engines, and examine their update transaction 
% time performance. 

% data models and structures that enable linking profiles from different database sources and 
% textual documents. 
% The data models support profiles having {multiple} values for the same attribute/relation as 
% well as {provenance} information; and the system accepts data from both structured and 
% unstructured sources. 

\iffalse
\cite{tagme-10}
\cite{dexter-mention2wiki-2013}
\cite{fr-tagme-to-WAT-2014}
\cite{reproduce-tagme}
\cite{gerbil-benchmk-framework-15WWW}
\cite{relation-as-latent-NN-Edinb-2018ACL}
\cite{foreseer-link-online-sources-2016BDSE}
\cite{IDEL-text-relational-mongo-NN-2018arxiv}
\cite{fr-AGDISTIS-to-MAG-graph-mutilingual-EL2017arxiv}
\cite{use-wiki-in-NEL-macqu2011WISE}
\cite{EREL-NR-with-coref-2017JIT}
\cite{Wikipedia-miner-2008cikm}
\cite{Singh17vldb-synthesize-rules}%by example
\cite{record-linkage-survey2006}
\cite{survey-entity-mention-2018}
\cite{survey-indexing-tech-2012}
\cite{comp-approx-blocking-ER-2016vldb}
\cite{evalu-entity-reso-approaches-2010vldb}
\cite{survey-similarity-meas-algo-sigmod06}
\cite{survey-entity-kb-shen2015}
\fi

\section{System Architecture \& Functions}\label{sec-arch}
This section covers the architecture of our entity linking system, and
outlines the functions of the components of the system. 
% This architecture has the capability to handle structured data, text data, and other data sources.
% Next, we present the physical implementation models we used to achieve best efficiency and effectiveness in the architecture. Finally, we describe the general linking principles used in the architecture for entity linking and resolution. 

\begin{figure}[h]
\center
\includegraphics[scale=0.52]{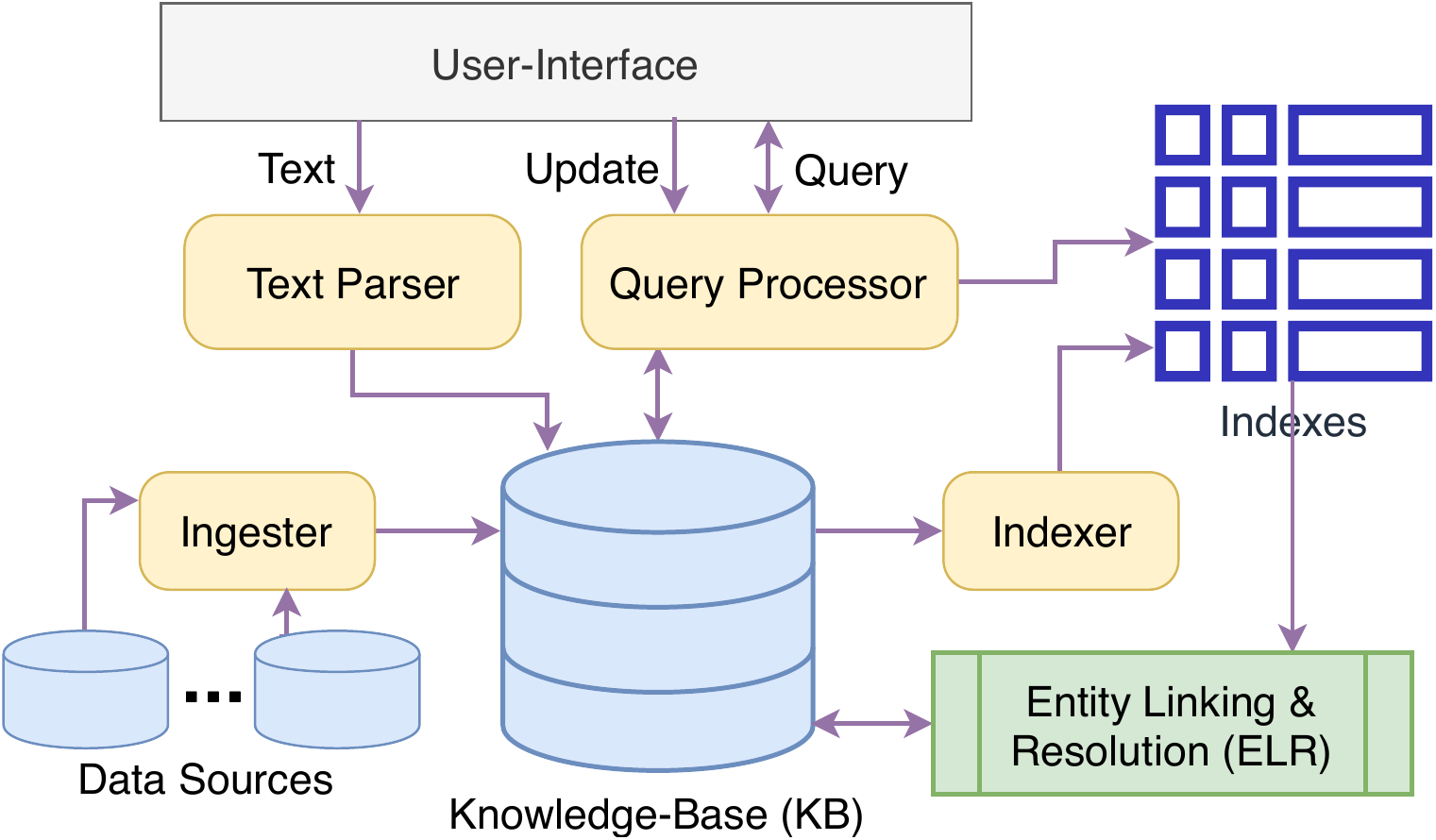}
\vspace{-8pt}
\caption{The architecture of our entity linking system}\label{fig-arch}
\vspace{-12pt}
\end{figure}

Figure \ref{fig-arch} presents an overview of the architecture of our system. 
Central in the system is the Knowledge-Base (KB) which is a graph of entity profiles
(details in Section~\ref{sec-models}). 
The profiles in the KB come from three sources: (a) ingested profiles from different 
data sources (through the {\it Ingester}) with no restriction on model; (b) extracted 
profiles from user-supplied textual documents (via the {\it Text Parser}); and 
(c) profiles created from the {\it User-Interface} (UI). 
The profiles are linked and indexed by the {\it Entity Linking \& Resolution} (ELR) 
and {\it Indexer} components respectively. And, all user interactions with the system 
are via the UI, mediated by the {\it Query Processor}.  

The following are brief details and functions of the components. 

\smallpara{The Ingester:}
maps entity descriptions from various data sources into graph-modelled profiles 
in the knowledge-base. Its operation is straight-forward and dependent on the 
respective models (or lack thereof) of the various sources of data.

\smallpara{The Text Parser:} reads textual user-inputs (e.g., documents, reports, etc.), and
extracts mentioned-entities and their relationships from the texts, and stores the extracted entity profiles into the knowledge-base. In our implementation, we use Stanford NER~\cite{stanfordNER}, Stanford POS tagger~\cite{stanfordPOS}, and Open IE 4.x~\cite{openIE4UW} for this purpose. 

The problem with the above-mentioned packages is that they may produce many triples 
(subject, relation, object) that do not reflect the original intention of authors in 
the writings. 
For example, extractions for the sentence ``{\it John said that, Peter has taken away the mobile phone}'', include the triple: ({\it ``Peter'', ``has taken away'', ``the mobile
phone''}). This extract is only syntactically correct. The semantic correctness of this extraction is, however, dependent on John's credibility/position. If John is a Police spokesman, for instance, then the chance of semantic-correctness would be high. However, 
if John is an adversary of Peter, for example, then the chance for the extraction to be 
correct would be low. 

Therefore, we developed some heuristic rules to filter ambiguous extractions. 
The rules: (a) replace coreferences (pronouns) with the actual entity-mentions; 
(b) remove extractions that are conditional, and indirect speech; and (c) filter 
extractions that describes feelings and emotions. The inputs to the rule-based 
filtering system (RbFS) are the text and labelling from Stanford NLP. The rules improve 
the F1-score of extractions by 18\% on average on our test datasets. The 
details of the RbFS is out of the scope of this paper. 

\smallpara{The Query Processor:} receives requests from the user-interface and 
responds based on the request type. An insert or update request is directly 
sent to the knowledge-base. For a query by profile identifier or keywords, 
the index is searched and then answers are retrieved from the knowledge-base.

\smallpara{The Indexer:}
 keeps the indexes up to date with the current system state. Whenever the knowledge-base 
 is updated, this component sends the update to the indexes. The indexes support
 users' queries and the ELR component. We use Elasticsearch,
 an open source distributed search and index engine, as 
 our index management system. Later on (in subsection~\ref{sec-index}), we 
 present details of the structure of the indexes in our system.  

\smallpara{The ELR component:}
 as the name suggests, is the main component in the system. 
 In principle, for every entity profile $p$ in the knowledge-base, indexes are 
 read for candidate matching profiles; calculations of the similarity of $p$ to
 each of the candidates are performed; and the knowledge-base is updated to store 
 the similarities. 
 A candidate $p'$ of $p$ from the indexes is a profile that is roughly similar 
 to $p$, i.e., the pair share some similar attribute and/or relationship values. 
 We remark that the fact that $p$ and $p'$ share a similar `word' does not 
 necessarily mean that they refer to the same real-world entity. 
 For instance, if $p$ is a {\tt male} with the name {\tt Pete} and $p'$ is a {\tt 
 female} and has a {\tt friend} called {\tt Peter}, then $p'$ can be a candidate 
 of $p$ as they share a similar value (i.e., {\tt Pete \& Peter}), but they do 
 not match. 
 Therefore, the indexes merely give a set of possible profiles that may match
 which require further evaluation. 

% Since indexes are crucial to the Linking component, we provide its details in the following subsection prior to describing the Linking component. 
%are two main components for entity linking, we describe them in details in the next two subsections. 

\section{Models}\label{sec-models}
In this section, we present the logical, physical, and indexing models used in 
our entity linking system.

\subsection{Modelling of Profiles}\label{sec-model-prof}
A real-world entity, naturally, has many attributes (or properties) and relates to 
other entities in multiple ways. An entity profile (simply, profile) captures some 
of the attributes and relationships of an entity; and another profile may 
capture different attributes and relationships of the same entity with possible 
overlaps and contradictions. 
For example, a person may have multiple profiles in the same/different sources 
of data (e.g., databases, knowledge-bases, social networking sites, etc.).

\smallpara{Entity profile structure.}
The data structure of profiles should be able to capture multiple values 
of the same attribute or relationship as well as their provenance information. 
This is because of the ever-changing or evolving nature of the properties and 
relationships of real-world entities. For example, a person may change 
his/her names, live at different addresses over time, have multiple marriages spanning 
different periods, etc. These changes lead to multiple values for attributes and relations 
and these values may be associated with provenance information. 

We represented an entity profile as a triple $p=\la id, {\calA, \calR}\ra$, 
where: $id$ is the identifier of the profile, ${\calA} =[a_1,\cdots, a_n]$ is a 
list of attribute-objects, ${\calR}=[r_1,\cdots, r_m]$ is a list of relationship-objects; 
and each $a\in \calA$, $r\in {\calR}$ is a set of ordered\footnote{attribute/relation
key-value pairs are first in the set, followed by the provenance data (if exists)} 
key-value pairs. 
Four profile examples are given in Table~\ref{exam-profile}, shown in our structure. 
Profile $p_1$ describes a {\tt person} entity: a {\tt male}  
called {\tt Peter} up to {\tt 1991} and now
called {\tt John}. He {\tt lived\_at} location {\tt L1} from {\tt 1989} to {\tt 1995}, 
{\tt owns} {\tt L1} since {\tt 1989} and has a {\tt friend} named {\tt Bob}. 
Our data structure for profiles is thus able to capture the multiplicity 
and provenance of values.

\begin{table*}[!t]
    \centering
    \caption{Example of profiles as triples}\label{exam-profile}
    % \small
    \vspace{-6pt}
\begin{tabular}{l|l}\hline
$p_1$ & $\la$ {\tt P1,} $\calA=$ {\tt [\{type: person\}, \{name: John\}, \{name: Peter, until: 1991\}, \{sex: m\}]},\\
          & $\quad  \calR=$ {\tt [\{lives\_at: L1, from: 1989, to: 1995\}, \{friend: P2\}, \{owns: L1\}]} $\ra$ \\\\ %\hline
$p_2$ & $\la$ {\tt P2,} $\calA=$ {\tt [\{type: person\}, \{name: Bob\}, \{name: John, until: 1990\}, \{bdate: 1980.12.12 \}]}, \\
          & $\quad  \calR=$ {\tt [\{lives\_at: L1, from: 1990, to: 2000\}, \{lives\_at: L2, from: 2001\}}, {\tt\{friend:P1\}]} $\ra$ \\ \\ %\hline
$p_3$ & $\la$ {\tt L1,} $\calA=$ {\tt [\{type: location\}, \{numb: 1\}, \{street: Brown Blvd.\}, \{post:2000\}]}, \\
          & $\quad  \calR=$ {\tt [\{owned\_by: P1, from: 1989\}]} $\ra$ \\\\
$p_4$ & $\la$ {\tt L2,} $\calA=$ {\tt [\{type: location\}, \{numb: 69\}, \{street: Brown Ave.\}, \{post:5000\}]} $\ra$ \\
         % & $\quad  \calR=$ {\tt [\{home\_of: P2, from: 2001\}]} $\ra$ \\
\hline
\end{tabular}
\end{table*}

\smallpara{Entity profiles graph.}
We use a graph model for modelling profiles, as it is capable of representing any 
number of attributes and relationships. Moreover, since $\calA$ and $\calR$ 
are defined as lists, instead of dictionaries, value multiplicity can be presented 
easily. Furthermore, the edges of the profiles graph allow the traversal of the profiles.

Formally, we use the following definition of an entity profiles graph, $G=(V,E,F_\calA)$, 
where: 
(i) $V$ is a finite set of nodes;
(ii) $E$ is a finite set of edges, given by $E \subseteq V \times V$; 
(iii) each node $v\in V$ (resp. edge $e\in E$) has a label $L(v)$ (resp. $L(e)$); and
(iv) each node $v \in V$ has an associated list $F_\calA(v) =[a_1, \cdots, a_n]$ of attribute-objects. 

\begin{figure}
    \centering
    \includegraphics[scale=0.76]{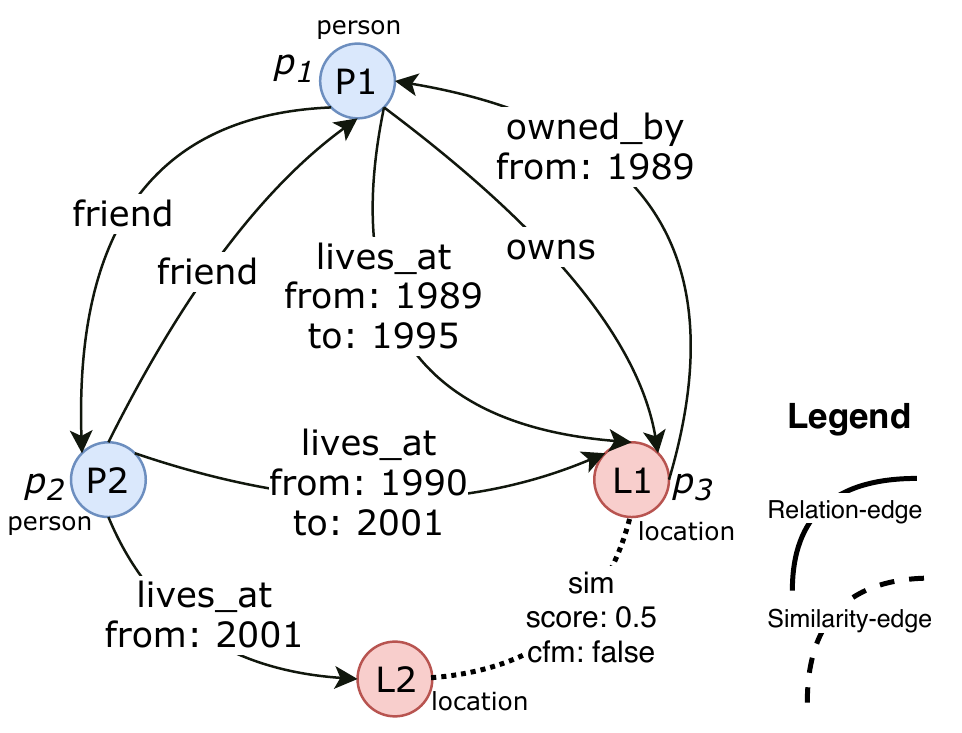}
    \vspace{-10pt}
    \caption{Graph representation of profiles in Table~\ref{exam-profile}}
    \label{fig:graph_example}
    \vspace{-12pt}
\end{figure}

A node in an entity profiles graph represents a profile $p$, identified by the profile 
$id$, and is associated with $\calA$ and $\calR$ as defined. Two types of edges exist in 
the graph.
One is called a  {\it relation-edge}, derived from $\calR$. That is, the edge $(rel, P1, P2)$ is an edge in the graph iff.: $P2$ is the value for relation $rel$ in $P1$ where $P1$ and $P2$ are profile/node identifiers. 
The second type of edge is called a {\it similarity-edge}, derived from the profile pair similarity and has the form $(sim,P1,P2,score,cfm)$ where $sim$ is a fixed label, $score$ is the similarity score (defined later in Section~\ref{sec-linking}), and $cfm$ is a binary indicator showing whether the link-state of a profile pair has been confirmed by a user. 
The indicator is necessary because, in sensitive systems like policing, we want  $100\%$ precision if two profiles are linked.  Thus, $cfm$ requires user-interaction (to be 
discussed further in Section~\ref{sec-linking}).   
Figure~\ref{fig:graph_example} is an example of the graph of profiles in Table~\ref{exam-profile}. 
Note that each node in the graph carries its attribute list (not shown in the diagram).

\subsection{Physical Model for Profiles}\label{sec-phy-model}
\begin{figure}[t]
	\centering
	\includegraphics[scale=0.3]{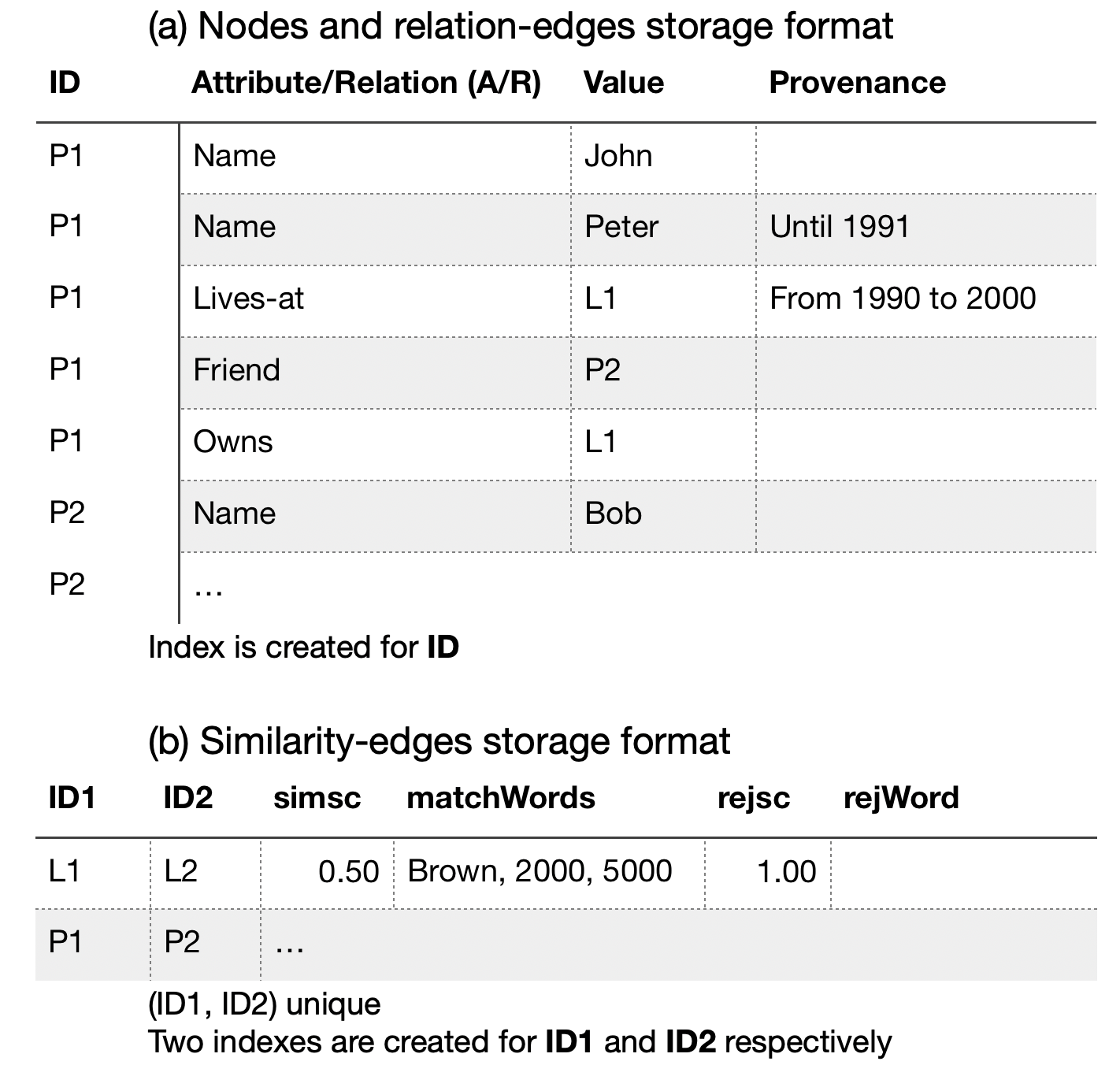}
	\vspace{-12pt}
	\caption{Physical model for storing the profiles graph}\label{tab-phy-model}
	\vspace{-13pt}
\end{figure}

Profiles are modelled as a graph; and the nodes and relation-edges are stored in one 
structure while the similarity-edges are stored in a separate structure. Similarity-edges 
are updated frequently as any profile change triggers a re-computation of similarities 
for the profile and other affected profiles. 
Therefore, storing similarity-edges in a separate structure 
improves the update efficiency. The two structures for storing the graph are called the 
{\it physical model} and shown in Figure \ref{tab-phy-model}.

The model in Figure \ref{tab-phy-model} is self-explainable; and the data in the two 
tables of the model are derived from some of the exemplar profiles in Table \ref{exam-profile}.
The table in Figure~\ref{tab-phy-model}(a) stores the nodes (profile ids and attributes) 
and relation-edges (relationships). Each node uses multiple lines and each line is for an attribute or relationship value pair with provenance details. 
The table in Figure~\ref{tab-phy-model}(b), on the other hand, is for the 
storage of similarity-edges 
and each pair of profiles has an entry in the structure. $simsc$ and $rejsc$ represent 
similarity score and rejecting score respectively (details in Section~\ref{sec-linking}).
Since the size of the table in Figure~
\ref{tab-phy-model}(b) is the square of the number of nodes/profiles, to reduce the size, 
a threshold may be used to filter out very lowly-scored entries. 

We realize that the performance of accessing the similarity-edge table plays a crucial 
role in the overall linking time performance due to its frequent update operations. 
Therefore, we show empirical results on three different implementation options of
the physical model in Section~\ref{sec-exp}. 

\subsection{Index Mappings}\label{sec-index}
Our aim is to design index structures to support users' search for profiles and 
support the candidate matching-profiles generation ({\em a.k.a} blocking) of the ELR component. 
Thus, we use Elasticsearch, a distributed index management system that can 
support multiple indexes with various structures.

Recall that, logically, each profile is a triple of the form $\la id, 
{\calA,\calR}\ra$ in the 
knowledge-base. We consider two options for building indexes for the profiles 
(discussed below), 
and both are configured with the double-metaphone phonetic analyzer~\cite{2-meta2}, 
and custom-built synonym and alias transformers.

\smallpara{Keyword search \& blocking indexes.}
The indexes that support keyword search of profiles and blocking for 
entity linking uses a set of words generated from the profiles. 
The set of words are 
values from the profile without provenance. That is, the provenance values, 
the structure, the attribute and relation names are all ignored, relationship 
targets (i.e., other profile ids) are replaced by the summary of the target, and 
all duplicate words are removed. 
For example, the target summary for $p_1$ in Table~\ref{exam-profile}, is a bag 
of the following values: ``{\tt John, Peter, m, 1, brown, 2000}''.

The `loose' structure of these indexes guarantee high recall of search
and blocking results.

\smallpara{Structured search indexes.}
The indexes for structural search consider the structure of the
profiles. For example, if a user wants to find a person with 
``$\tt name:John,lives\_at:1\ Brown\ street-\{until: 2000\}$'', the index should 
enable $p_1$ in Table \ref{exam-profile} to be found. To support such structural 
search, we build indexes with the {\it nested mappings} in Elasticsearch for 
profiles structured as JSON objects with:
% \begin{align}
\[
\small
 {\calA} = {\tt [\{A_1:v_1, from:t_0,to:t_1\},\{A_1:v_2,to:t_2\},\{A_2:v_3\},\cdots]}, \label{eq-index-array}
\]
% \end{align}
where $t_0,t_1,t_2$ are date time values; $\calR$ mapping similarly defined.

We remark that the usage of nested mappings is critical to the preservation of the
correct semantics of the multiplicity of values and their associated provenance 
information in Elasticsearch. 
Otherwise, Elasticsearch indexes the profiles in a `flat-format' 
of the form: {$\tt \{A_1:[v_1,v_2],A_2:[v_3],from:[t_0], to:[t_1,t_2] \}$} -- which 
loses semantics and leads to errors and very low precision. 

In settings where smaller and more precise blocking are required, the 
nested-mapped indexes should be used.

\section{Linking of Profiles}\label{sec-linking}
This section presents a description of our profiles 
comparison and linking processes. First, we highlight some 
relevant preprocessing steps. Then, we detail the profile-pair
comparison and evaluation; and finally, give a brief overview of the match prediction and confirmation.

\smallpara{Preprocesses.}
Prior to the calculation of the similarity between profile pairs, some preprocessing are 
necessary. For example, consider person and location entities: it is important to tackle 
the disparate representation of the same names and addresses respectively. The name 
{\tt Richard} is often aliased as {\tt Dick}; and the street-type {\tt Boulevard} is often 
shortened as {\tt BLVD}. To enable Dick to match Richard, a dictionary of name aliases of 
people is created (similarly, for addresses). Each name/address in a profile is checked against 
the dictionary. If the name has an alias, the name is expanded in the form ``name alias'', e.g., 
``Richard Dick''. Similar operations are performed on the initials, and pre-/post-fixes 
of names. 

\begin{figure*}[t]
    \centering
    \includegraphics[scale=0.39]{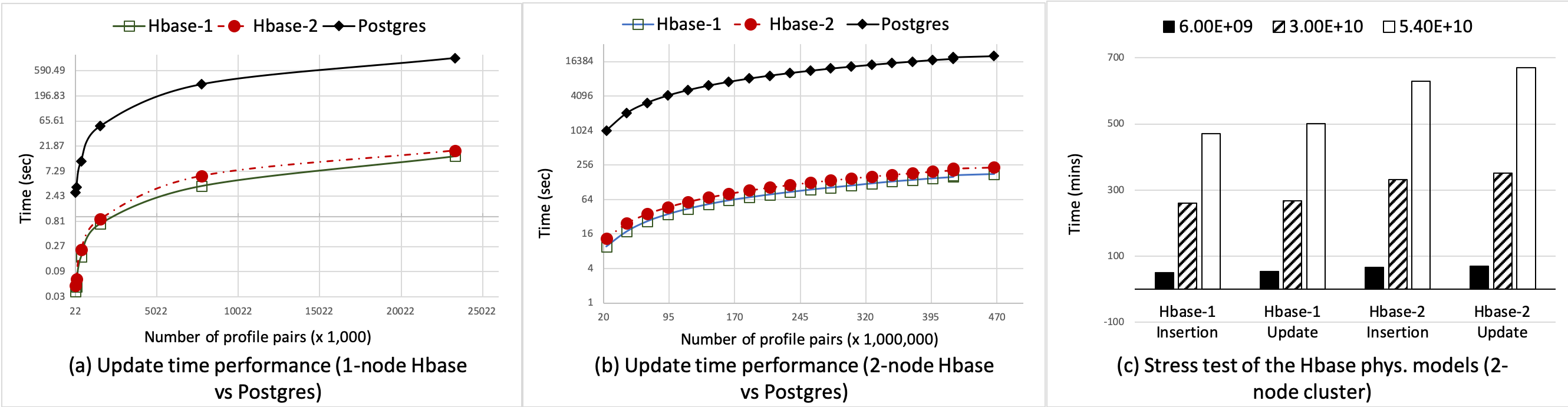}
    \caption{Time performance of update transactions}
    \label{fig:performance}
\end{figure*}{}

\smallpara{Similarity evaluation.}
Given two profiles $p_1$ and $p_2$, our entity linking method uses two scoring and one decision 
processes to determine whether they refer to the same entity in the real-world. The two scores 
are the similarity score, $simsc$, and the rejection score, $rejsc$; while the decision process 
is a data-dependency-based prediction model. We discuss the scoring here.

Given a profile $p$, we use the notation $X \in p$ to represent either an attribute $X$ in 
$p[\calA]$ or a relation $X$ in $p[{\calR}]$. 
The similarity score, $simsc$, of two profiles $p_1, p_2$, is calculated as follows:
\[
simsc(p_1,p_2) = \sum_{X \in p_1, p_2} {\calM}(p_1.X,\ p_2.X)\cdot {\calI}_X(p_1.X, p_2.X),
\]
where ${\calM}$ is a function that returns a value indicating the level of approximate match 
between a pair of values for the same attribute/relation $X$, and ${\calI}$ returns the level of
information supplied by the match ${\calM}$ for the values of $X$. 

The function ${\calM}$ considers many factors, dependent on the attribute/relation; 
and the values of an attribute/relation are in the form of a bag of words after 
synonym/alias expansion with provenance data. For example, to evaluate a {\tt name} 
match for person entities, ${\calM}$ considers the initials, ordering, post-/prefixes, 
aliases, and phonetics of names, as well as n-gram matching of character/word sequences. 
Edit distance is used after n-gram matching to improve accuracy and efficiency. 
If two values match within a user-specified threshold, then the provenance 
information are considered. 

\sloppy
The function ${\calI}$ returns the highest {\em information level} of matching values. 
For example, for the name-pair ``{\tt John Smith White}'' and ``{\tt Jones Smiths Green}'', 
the ${\calI}_{name}$-weight is derived as: 
\[max\{\frac{inf(John)+inf(Jones)}{2}, \frac{inf(Smith)+inf(Smiths)}{2}\}.\]
The function $inf(w)$ indicates the probability of two profiles to be linked if they match 
on the value $w$. Note that, in this example, the name-pair ``Green'' and ``White'' are not 
considered in the evaluation of $\calI$ as they are dissimilar (i.e., have low $\calM$ value).
Intuitively, if a word $w$ is rare, it has high $inf(w)$-value. 
Consider the two first-names `John' and `Cherith'. 
When two profiles share the name `Cherith', the probability for the two profiles to be 
linked is much higher than when two profiles share the name `John'.

The $inf(w)$-value of a word $w$ is controlled by two factors: the number $m(w)$ of profiles 
sharing the word, and the number $k(w)$ of real-world entities shared by the profiles sharing 
the word. If $k(w)$ is large, the fact that the $m(w)$ profiles share the same word contributes 
very little to the linking, and $inf(w)$ should be small. 
When the total number of profiles increases, the chance for two profiles to share the same word becomes larger but the crucial control of the probability is still by $k/m$. 
That is, $inf(w) \varpropto m/k$; and $m$ can be easily obtained from word statistics 
but $k$ is 
often unknown. A large $m$ does not mean a large $m/k$ ratio. 
We use a variation of the Sigmoid function to estimate $inf(w)$, given as: \[inf(w)=\frac{1}{1+exp(\alpha\cdot m(w)-\beta)},\] 
where $\alpha,\beta$ control the steepness of the decay curve and its mid-point 
respectively. 
For a given $w$, $k(w)$ can be empirically estimated and linked to $\beta$.
However, in general, our empirical results suggest $\alpha = 0.1$ and $\beta=60$ 
are suitable settings for our applications.

The $rejsc$ score, on the other hand, is based on a simple penalty 
system. Given two profiles with a high overall $simsc$ score, a penalty of $1$ is
added to their $rejsc$ score if the pair are dissimilar on a {\em key} attribute/relation (determined by application and domain). 
For example, in a law enforcement context, one such key attribute for person and location
entities is {\tt birth-date} and {\tt zip-code} respectively.

\begin{figure}[t]
	\centering
	\includegraphics[scale=0.38]{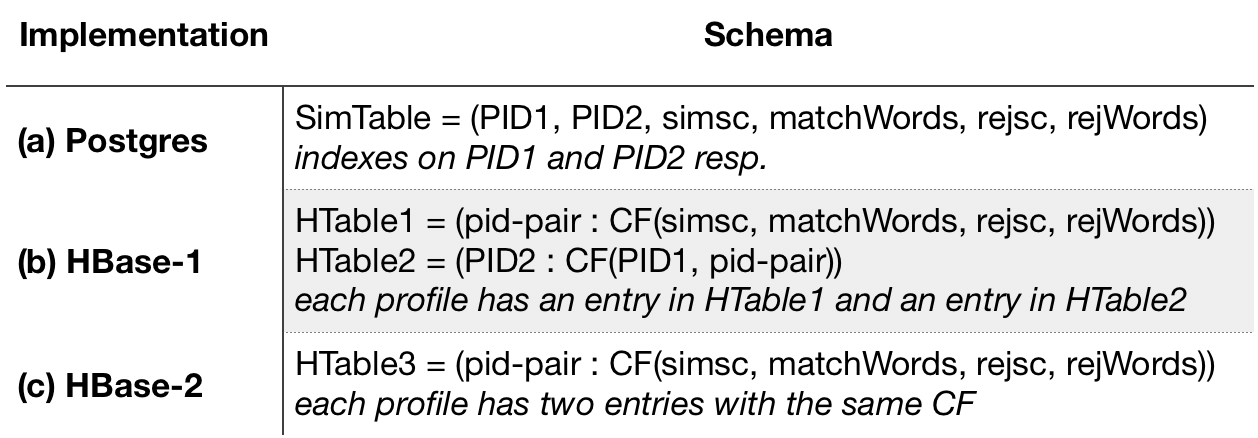}
	\vspace{-12pt}
	\caption{Implementation options of the similarity structure}\label{tab-impl-options}
	\vspace{-10pt}
\end{figure}

\smallpara{Match prediction \& confirmation.}
As mentioned earlier, we use a data-dependency aided decision 
model to predict whether a given pair of similar profiles refer to the
same real-world entity. This decision model is a major topic (and we refer interested readers to the paper on it in~\cite{KwashieLLLSY19}). The approach eliminates the challenge of and need for fine-tuning dis/similarity thresholds for 
approximate matching, through the use of a discovery algorithm that 
learns matching rules in labeled data. 
The match prediction model achieves high precision without significant
compromise of recall. 

In some applications, even accurate prediction of 
the linked status of two profiles require human confirmations. Thus, 
our entity linking system supports this scenario, allowing the keeping
of domain experts in the loop. Indeed, every similarity-edge between profiles 
carry the data structure for the confirmation of predicted matches (when needed). For example, in Figure~\ref{fig:graph_example}, the similarity-edge between 
nodes {\tt L1} and {\tt L2} is not confirmed (i.e., {\tt cfm: false}).

\section{Experiments}\label{sec-exp}
In this section, we empirically evaluate the performance of the three
%show the results of experiments on 
different implementations of the physical model. We remark that, the 
accuracy of the ELR system is already evaluated in~\cite{KwashieLLLSY19}.

We note that updating the pairwise similarities of profiles is a major 
performance bottleneck. This is because, for every 1,000 profiles, the 
updated similarity entries are around 20,000-100,000. 
Therefore, we examine the time efficiency of accessing the similarity 
structure (Figure \ref{tab-phy-model}(b)).

All procedures in the work are implemented in Java, and the entity 
linking system runs on Ubuntu 18.04 machine(s). For single-machine tests,
the experiments were run on an Intel(R) Core(TM) i7-7700 CPU @ 3.60GHz
computer with 32GB of memory. In the cases where multi-node HBase
clusters are required, an Intel(R) Core(TM) i7 CPU @ 2.30GHz
computer with 16GB of memory is added. The versions of Postgres and
HBase used are 9.6.12 and 1.4.8 respectively.

\subsection*{Efficiency of Similarity Storage Structure}
We present our experiment results on the efficiency of accessing the similarity structure 
on different platforms with different implementations. We tested the implementation in
Postgres\footnote{It is noteworthy that the performance difference of the Postgres
implementation for ON/OFF AUTOCOMMIT settings is marginal. Thus, we report the best 
(i.e., AUTOCOMMIT OFF).} and HBase, with schemas summarized in Figure~\ref{tab-impl-options}.

The operations to access the similarity structure include search, insertion, update and deletion. Since the similarity is for a pair, the search must be supported from either ID. 
We created two indexes for this purpose in the relational option of Postgres. 
With the HBase options, CF means a column family which is a dictionary of key-value pairs with the keys listed in the brackets. The `id-pair' is constructed by ID1+"-"+ID2. 
In the case of HTable3 in Figure~\ref{tab-impl-options}(c), the second id-pair is ID2+"-"+ID1.

\smallpara{Performance of the physical model on small to large data.}
In this experiment, we examine the relative update transaction (involving search, insert \& 
delete operations) time performance of the three physical model implementations 
(in Figure~\ref{tab-impl-options}) over small to large datasets. The results for:
(a) small- to medium-sized data (i.e., $23K$ to $23M$ profile pairs), and (b) medium- to large-sized data (i.e., $23M$ to $468M$ profile pairs) are presented in
Figure~\ref{fig:performance} (a) \& (b) respectively.
%In Figure~\ref{fig:performance}(a) presents the relative time performance of the three physical implementation options in Figure~\ref{tab-impl-options}. 
The $x$-axes show the number of profile pairs updated; and 
the $y$-axes give the average time, in seconds (on a log2 scale), taken to perform update transactions (over five iterations).

For case (a) above, the HBase-1 and HBase-2 implementations are on a single-node cluster 
for a fair comparison with the Postgres implementation; and for case (b), the HBase implementations are on a two-node cluster. 
The results show that in all cases, of the three implementation options, the relational option (Postgres) is significantly slower than the HBase counterparts; and the HBase-1
implementation (i.e., option (b) in Figure~\ref{tab-impl-options}) is the better of 
the two HBase options. It is also noteworthy that there is no significant performance 
difference between the HBase implementations on single-node and two-node clusters.

\smallpara{Stress test of HBase implementations.}
In this experiment, we perform further tests to examine the insertion and update (replacement) operations of the best-performing models (i.e., the two HBase models). 
We consider three data sizes: $6$, $30$, and  $54$ billion profile pairs. As the 
results in Figure~\ref{fig:performance}(c) show, the insertion operations are, 
as expected, more efficient than update operations for both models over the 
three datasets. Moreover, both the insertion and update operations are scalable 
for both implementations on very small-sized (i.e., just a two-node) cluster.

\section{Conclusion}
In this paper, we present the details of the entity linking system that
powers our entity linking method called \certus. 
We describe the architecture of the system, the graph and data models, and
index structures used to support the multiplicity and provenance of attribute 
and relation values, for effective entity linking and resolution. 
Further, we give the details of the physical model for storing the entity 
profiles graph, and discuss three different implementations of the structure
for storing the similarity-edges. Due to the frequency of the update transaction
of similarity-edges, we perform experiments to evaluate the time performance
of accessing the similarity structure on two
state-of-the-art database management systems (HBase and Postgres) to demonstrate the relative performances of the three different implementations. 
The empirical results show a generally good performance for all implementation 
options. In particular, the HBase implementation options, with even just one- or 
two-node clusters, scale very well for huge data sizes.

\begin{acks}
This work is supported by Data to Decisions CRC, ILE/ELR Project DC160031, funded by
the Australian Commonwealth Government's CRC Programme. And, partly supported by Data61 - CSIRO.
\end{acks}

\balance

\bibliographystyle{ACM-Reference-Format}
\bibliography{sample-authordraft}

\end{document}